\documentclass[11pt,a4paper]{article}
\usepackage{jcappub}
  
\title{The High-$z$ Quasar Hubble Diagram}
\author[1]{Fulvio~Melia%
\note{John Woodruff Simpson Fellow.}}
\affiliation{Department of Physics, the Applied Math Program, and Steward Observatory \\
              The University of Arizona \\
              Tucson, AZ 85721}
\emailAdd{fmelia@email.arizona.edu}

\abstract{Two recent discoveries have made it possible for us to begin using high-$z$ quasars as
standard candles to construct a Hubble Diagram (HD) at $z> 6$. These are (1) the 
recognition from reverberation mapping that a relationship exists between the 
optical/UV luminosity and the distance of line-emitting gas from the central ionizing 
source. Thus, together with a measurement of the velocity of the line-emitting
gas, e.g., via  the width of BLR lines, such as Mg II, a single observation can 
therefore in principle provide a determination of the black hole's mass; 
and (2) the identification of quasar ULAS J1120+0641 at $z=7.085$, which has 
significantly extended the redshift range of these sources, providing essential 
leverage when fitting theoretical luminosity distances to the data. In this paper, 
we use the observed fluxes and Mg II line-widths of these sources to show
that one may reasonably test the predicted high-$z$ distance versus redshift
relationship, and we assemble a sample of 20 currently available high-$z$ quasars 
for this exercise. We find a good match between theory and observations,
suggesting that a more complete, high-quality survey may indeed eventually 
produce an HD to complement the highly-detailed study already underway (e.g.,
with Type Ia SNe, GRBs, and cosmic chronometers) at lower redshifts. With
the modest sample we have here, we show that the $R_{\rm h}=ct$
Universe and $\Lambda$CDM both fit the data quite well, though the smaller
number of free parameters in the former produces a more favorable outcome
when we calculate likelihoods using the Akaike, Kullback, and Bayes
Information Criteria. These three statistical tools result in similar probabilities,
indicating that the $R_{\rm h}=ct$ Universe is more likely than $\Lambda$CDM
to be correct, by a ratio of about $85\%$ to $15\%$.
}

\begin{document}
\maketitle
 
 \flushbottom


\section{Introduction}
A powerful method of probing the cosmological expansion involves the acquisition
of distance versus redshift data for sources whose absolute luminosity is accurately
known. Plotting this information to produce (what is commonly referred to as) the Hubble
Diagram (HD) then provides us with the expansion history of the Universe, and since
the cosmic evolution depends critically on its constituents, measuring distances over
a broad range of redshifts can in principle place meaningful constraints on
assumed cosmological models. As is well known by now, it was this program
that lead to the discovery of dark energy through the use of Type Ia supernovae
\cite{R98,P98,P99,G98,S98}.
These events produce a relatively well-known luminosity, permitting
them to function as reasonable standard candles, under the assumption that the 
power of both near and distant explosions can be standardized with the same 
luminosity versus color and light-curve shape relationships.

However, being reasonably sure that something other than (luminous and cold dark)
matter and radiation must be present in the Universe is a far cry from
understanding what dark energy is, or even knowing what its equation 
of state $p_{de}=w_{de}\rho_{de}$ must be, in terms of its pressure $p_{de}$,
its energy density $\rho_{de}$, and the dimensionless parameter $w_{de}$
that may or may not be changing with time. The standard model of cosmology 
($\Lambda$CDM)  posits that $w_{de}=-1$ at all times, the simplest assumption 
one can make based on Einstein's cosmological constant $\Lambda$. This form
of $\rho_{de}$ may be a manifestation of vacuum energy, though its value
would be at odds with the prediction from quantum mechanics. 

But as impressive as the use of Type Ia SNe has been, several important 
limitations mitigate the overall impact of this work. Principal among these is
the fact that even excellent space-based platforms such as SNAP 
have difficulty observing these events at redshifts $> 1.8$. 
Since much of the interesting physics driving the evolution of the Universe
occurred well before this epoch, we are therefore quite restricted in what 
we can learn from Type Ia SNe alone.

In addition, an incompatibility is now emerging between the use of the standard 
model to interpret Type Ia SNe and its application to other equally important 
observations, such as those of the cosmic chronometers \cite{MM13} 
and the cosmic microwave background (CMB) \cite{M12a,M12b}. 
Growing tension between these measurements and the predictions of $\Lambda$CDM 
suggest that the standard model may not be providing an accurate 
representation of the cosmological expansion at high redshifts ($z>>2$). 
For example, the Wilkinson Microwave Anisotropy Probe (WMAP) 
\cite{B03} and {\it Planck} \cite{Ade13} have uncovered 
several anomalies in the full CMB sky that appear to indicate possible 
new physics driving the growth of density fluctuations in the early Universe. 
These include an unusually low power at the largest scales and an apparent 
mutual alignment of the quadrupole and octopole moments, for which there 
appears to be no statistically significant correlation in $\Lambda$CDM. Their 
combined statistical significance is therefore equal to the product of their 
individual significances, suggesting that the simultaneous observation in 
the context of the standard model of the missing large-angle correlations 
with probability $<0.1\%$ and a low-$l$ multipole alignment with probability 
$\sim 4.9\%$ is likely at the $<0.005\%$ level \cite{C09}.

However, even 
at low redshifts, there are limitations to how well the Type Ia supernova
data can be interpreted with an empirical model such as $\Lambda$CDM, 
because the data cannot be determined independently of the assumed 
cosmology---the supernova luminosities must be evaluated by optimizing 
at least 4 parameters simultaneously with those in the adopted model. This 
renders the data compliant to the underlying theory, so the model-dependent 
data reduction cannot be ignored in any comparative analysis between competing 
cosmologies. For example, we recently demonstrated that the Type Ia supernova
HD in the best fit $\Lambda$CDM model is virtually indistinguishable from that in
the $R_{\rm h}=ct$ Universe \cite{M07,M12b}, over a redshift range 
extending all the way to $z\sim 6$ and beyond \cite{M12a}. 

Having said this, there do exist model-independent data, such as the
so-called cosmic chronometers \cite{J02}, that one can use to
test different expansion scenarios at low redshifts. In this approach, luminous 
red galaxies provide us with a method of measuring the universal expansion 
rate $H(z)$ in a model-independent way. However, our recent
analysis of these data, comparing the standard model with the
$R_{\rm h}=ct$ Universe (discussed more extensively below),
has shown that model selection tools, such as the Akaike, Kullback, and Bayes
Information Criteria, disfavor $\Lambda$CDM \cite{MM13}.
On the basis of these data, the likelihood that the standard model is closer
than $R_{\rm h}=ct$ to the correct cosmology is less than $\sim 8-18\%$
(depending on which criterion one uses).

These are some of the reasons for seeking other kinds of standard candle
to extend the HD to redshifts well beyond the Type Ia SNe range. An example
of such a category of sources currently being studied for this purpose are gamma-ray
bursts (GRBs) \cite{G04,S07,Q12} which, like Type Ia SNe, are transient and believed 
to result from explosive stellar deaths (though in this case with a mass much bigger 
than that of the Sun). Recently, the {\it Swift} spacecraft has added considerably 
to the GRB database, from $z\sim 0$ all the way to $z\sim 6$. We ourselves have
compiled an updated HD using GRBs detected in recent years, providing
a means of testing cosmological models at intermediate redshifts, between 
the Type Ia SN range and $z> 6$ \cite{W13}. And here also we
found that the standard model does not appear to be preferred by the 
observations. A comparative analysis between $\Lambda$CDM and the
$R_{\rm h}=ct$ Universe using the GRB data and the Akaike, Kullback,
and Bayes Information Criteria suggests that the likelihood of $\Lambda$CDM
being the correct cosmology instead of $R_{\rm h}=ct$ is only
$\sim 4-15\%$.

In this paper, we highlight several key recent discoveries
that now allow us to suggest the use of high-$z$ quasars as standard 
candles to construct a Hubble Diagram at redshifts beyond $\sim 6$, but 
only under fairly stringent conditions. The first of these novel results  is 
that the Mg II FWHM and UV luminosity of quasars beyond $z\sim 6$ appear
to be correlated. Since reverberation mapping of their broad lines also reveals 
a relationship between the distance of the line-emitting gas from the central 
ionizing source and the optical/UV flux, these two features together can 
therefore yield a possibly useful measurement of the black-hole 
mass. In addition, estimates of their bolometric power using the 
$F_{3000}$ flux density inferred from their fitted continuum suggest
that the most luminous quasars at $z> 6$ may be accreting near 
their Eddington limit, $L_{\rm Ed}$ \cite{W10,D11}. More importantly for 
the analysis we will carry out in this paper, the observed range of Eddington 
factors, $\lambda_{\rm Ed}$, appears to be narrowing as $z\rightarrow 6-7$, 
centered on a value close to one. Here, $\lambda_{\rm Ed} \equiv L_{bol}/
L_{\rm Ed}$, in terms of the bolometric ($L_{bol}$) and Eddington 
($L_{\rm Ed}$) luminosities. Thus, knowledge of their redshift and UV
spectrum makes them potentially viable sources to use in order to 
construct a Hubble Diagram.

The second significant discovery that makes
this idea viable was the detection of a luminous quasar at redshift 
$z=7.085$ \cite{M11}. As we shall see, by extending the 
redshift coverage from the previous record around 6.4 to over 7, 
this single event has greatly improved the leverage attainable
when fitting theoretical luminosity distances to the data. This
is especially true in view of the growing realization that quasars
tend to accrete closer to $L_{\rm Ed}$ as their redshift increases
(see, e.g., ref. \cite{Shen12}). At the very least, there appears to 
be a transition from sub-Eddington to near Eddington-limited accretion 
as the redshift increases past $\sim 6$ \cite{W10,D11}, though
this inference may be due in part to selection effects, since it is
primarily based on the observation of the most luminous sources
at this redshift. As we shall see in subsequent sections, the inclusion
of fainter quasars may somewhat mitigate this perceived general trend.

Any attempt at using supermassive black holes as standard
candles comes with several important caveats, so the kind of analysis 
we are conducting here should not be viewed in isolation. The true 
benefit of this work will emerge only when the results are compared 
to efforts using Type Ia SNe at lower redshifts and, eventually,
to the Hubble Diagram constructed from GRBs at intermediate
redshifts. We will discuss some of the more obvious caveats in
\S~2 below, and then demonstrate how high-$z$ quasars may
be used to construct an HD in \S~3. We will then compare this
HD with two cosmological models in \S~4, and discuss the results 
and present some conclusions in the final section of the paper.

\section{High-$z$ Quasars as Standard Candles}
Reverberation mapping of the broad-line region in quasars produces
a tight relationship between the distance $R$ of the line-emitting gas
from the central ionizing source, and the optical/UV luminosity, $L_{UV}$
\cite{B82}. The form of this dependence,
\begin{equation}
R\propto L_{UV}^{0.5}\;,
\end{equation}
is consistent with straightforward ionization models \cite{K00,B09}. 
Thus, the simultaneous measurement of the quasar's luminosity and 
the velocity of its line-emitting gas, e.g., via the observation of its 
Doppler-broadened Mg II line, is sufficient, in principle, to determine 
the gravitational mass $M$ of the central supermassive black hole \cite{W99}.
However, one must be aware of the various sources of uncertainty still
associated with these measurements, which limit the accuracy of the
black-hole mass determination to $\approx 0.4-0.5$ dex
\cite{S08}. Claims have been made that the accuracy may be as
good as $\approx 0.3$ dex \cite{S10}, though these may be unrealistic
(see also refs. \cite{Shen13,P10,P13} for a review of the reliability 
and accuracy of this method).

This limited uncertainty is important because the use of 
high-$z$ quasars as standard candles relies quite critically on
how accurately $M$ can be determined. If high-quality
line and continuum measurements are available, one can use the
relationship \cite{V09}
\begin{equation}
\log M=6.86+2\log{{\rm FWHM(Mg II)}\over 1,000\;{\rm km\hskip 0.05in s}^{-1}}+
0.5\log{L_{3000}\over 10^{44}\,{\rm ergs\hskip 0.05in s}^{-1}},
\end{equation}
in terms of the Mg II line width, FWHM(Mg II), and the luminosity
$L_{3000}$ at rest-frame $3000\;\AA$. This mass-scaling
relationship was obtained using several thousand high-quality
spectra from the SDSS DR3 quasar sample \cite{Sc05}, 
with a calibration to the H$\beta$ and C IV relations. 
The scaling law was applied to the subset of the DR3
quasar sample used to establish the luminosity \cite{R06b}
and black-hole mass \cite{V08} functions. Equation~(2.2) has 
been employed quite effectively to measure quasar masses
\cite{W10} in the analysis of nine Canada-France 
High-$z$ Quasar Survey (CFHQS) sources, and an additional eight SDSS 
sources with near-IR Mg II spectroscopy of sufficient quality to match 
that of the CFHQS sample. The SDSS quasars were originally reported in 
refs. \cite{J07,K07,K09}. All 17 of these sources, together with several 
others from ref. \cite{D11}, and the newest quasar ULAS J1120+0641 
at $z=7.085$ \cite{M11}, are included in Table 1 below. 

The $F_{3000}$ flux density is measurable to an accuracy
of about $10\%$ \cite{W10}. The 
FWHM is measurable to a corresponding accuracy of about 
$15\%$. Thus, determining the black hole masses by inserting 
the extreme values of $L_{3000}$ and the FWHM (based on 
their rms uncertainties) into the above equation yields a mass 
estimate uncertain by a factor of several (i.e., the aforementioned 
$\approx 0.4-0.5$ dex; \cite{Shen13,P13}). This is evident from the 
range of masses quoted for each source listed in Table 1. As we
shall show below, the measured values of FWHM(MgII) and $L_{3000}$ 
may be used for tests of cosmological models without actually calculating
the black-hole mass $M$ itself, though the uncertainty in these quantities
carries through to a determination of the HD constructed
from them.

The luminosities and masses inferred from the
measured fluxes, and the luminosity distance inferred from the observed
redshift, all depend on the assumed cosmology. Fortunately, we will
not need to use these inferred quantities to construct the HD, but
show their values here for illustrative purposes. The entries listed in 
Table 1 correspond to a $\Lambda$CDM model
with a Hubble constant $H_0=70$ km s$^{-1}$ Mpc$^{-1}$, and a scaled
matter density $\Omega_m=0.28$, where $\Omega_m\equiv\rho_m/
\rho_c$. The critical density $\rho_c\equiv 3c^2H_0^2/(8\pi G)$ is
determined under the assumption that the Universe is flat, so the total 
scaled energy density $\Omega\equiv\Omega_m+\Omega_r+
\Omega_\Lambda$ equals 1. The other quantities in this expression
are the corresponding radiation ($\Omega_r\equiv\rho_r/\rho_c$) and 
dark energy ($\Omega_\Lambda\equiv\rho_\Lambda/\rho_c$) densities.

The values of $F_{3000}$ were derived from the fitted continuum, and 
their uncertainties include $10\%$ added in quadrature to account for 
the absolute flux calibration uncertainty. The monochromatic luminosity
is only a fraction of the total power produced by the quasar, so a bolometric 
correction $\eta$ must be applied to find its total luminosity, $L_{bol}$ 
($\equiv \eta L_{3000}$; this is shown for $\Lambda$CDM in the fifth 
column of Table 1). These values were obtained from $L_{3000}$ using 
a bolometric factor $\eta=6.0$ \cite{R06a,J06}, though the
estimation of $L_{bol}$ from a single monochromatic luminosity
can be quite uncertain for individual objects, given the diversity
of quasar spectral energy distributions (SEDs) (see, e.g., the cautionary
discussions in \cite{R06a}).

The SEDs in ref. \cite{R06a} update
the mean SED from ref. \cite{E94}, often used previously to derive
bolometric luminosities and accretion rates. These newer SEDs were constructed
from 259 SDSS quasars, combining SDSS magnitudes and {\it Sptizer} IRAC flux
densities, though with some ``gap repair" in other bands for which some
sources have no measurements. The quasar spectra were also corrected for
host galaxy contamination, using scaling relationships among host
bulge luminosity, bulge mass, black-hole mass, and Eddington luminosity,
to estimate the contribution of host galaxy light to the quasar SEDs 
\cite{D03, V06}. The quasar luminosity versus host luminosity 
relationship at optical frequencies provides a reasonable estimate 
of the host galaxy contribution, under the assumption that the 
quasars are emitting at their Eddington limit.

An important caveat with this work is that in order to construct mean
SEDs, the flux densities of each individual object can be compared
or combined with those of other quasars in the sample by adopting
a particular cosmology. Thus, the process of obtaining an average
value of $\eta$ is not entirely free of the presumed background
expansion scenario. Insofar as comparing $R_{\rm h}=ct$ with
$\Lambda$CDM using the high-$z$ quasar sample is concerned, 
this is not a serious problem because, as we shall see, the
concordance $\Lambda$CDM model essentially replicates the
dynamics of $R_{\rm h}=ct$, so that if one were to use the
latter to construct the average quasar SED \cite{R06a,R06b},
the outcome would be very close
to what they obtained using a standard flat cosmology with
$H_0=70$ km s$^{-1}$ Mpc$^{-1}$, $\Omega_m=0.3$,
and  $\Omega_\Lambda=0.7$.

\begin{table*}
\center
  \centerline{Table 1. High-$z$ Quasars}\vskip 0.1in
  \begin{tabular}{lccccrl}
\hline\hline
&&&&&& \\
    Name & $\quad z\quad$ & $M^\dag$ & FWHM (Mg II) & $L_{bol}^\dag$ & $d_L^{\Lambda{\rm CDM}\;\dag}$ & Ref. \\
              &  &$10^8\;M_\odot$ & km s$^{-1}$ & $10^{45}$ ergs s$^{-1}$&Glyr& \\ \hline
&&&&&& \\
ULAS J1120+0641& 7.085 &13--35&$3800\pm200$&252&227.21&\cite{M11}\\
CFHQS J0210-0456&6.438&0.4--1.35&$1300\pm350$&22--28&203.32&\cite{W10}\\
SDSS J1148+5251& 6.419 &44--87 &$6000\pm850$ &360  &202.62&\cite{W10}\\
CFHQS J2329-0301&6.417&2.1--2.9&$2020\pm110$&37--47&202.55&\cite{W10}\\
SDSS J1030+0524& 6.310 &12--24&$3600\pm100$&180&198.63&\cite{K07}\\
CFHQS J0050+3445&6.253&22--31&$4360\pm270$&185--226&196.54&\cite{W10}\\
SDSS J1623+3112& 6.250 &11--21&$3600\pm411$&171&196.43&\cite{J07}\\
SDSS J1048 + 4637&6.198&25--62&$3366\pm532$&304&194.53&\cite{D11}\\
CFHQS J0221-0802&6.161&2.3--14.5&$3680\pm1500$&27--33&193.18&\cite{W10}\\
CFHQS J2229+1457&6.152&0.7--1.9&$1440\pm330$&32--40&192.85&\cite{W10}\\
CFHQS J1509-1749&6.121&27--33&$4420\pm130$&238--290&191.72&\cite{W10}\\
CFHQS J2100-1715&6.087&6.9--12.3&$3610\pm420$&53--65&190.47&\cite{W10}\\
SDSS J0303-0019& 6.080 &2.6--5&$2300\pm125$&53&190.22&\cite{K09}\\
SDSS J0353 + 0104&6.072&9--22&$3682\pm281$&146&189.93&\cite{D11}\\
SDSS J0842 + 1218&6.069 &11--27&$3931\pm257$&155&189.82&\cite{D11}\\
SDSS J1630 + 4012&6.058&6--14&$3366\pm533$&94&189.42&\cite{D11}\\
CFHQS J1641+3755&6.047&1.6--3.4&$1740\pm190$&64--80&189.02&\cite{W10}\\
SDSS J1306 + 0356&6.020&19--36&$4500\pm160$&192&188.03&\cite{K07}\\
CFHQS J0055+0146&5.983&1.7--3.3&$2040\pm280$&34--42&186.68&\cite{W10}\\
SDSS J1411+1217&5.950&6--10&$2400\pm150$&240&185.48&\cite{K07}\\ 
&&&&&& \\ \hline\hline
  \end{tabular}
{\vskip 0.1in $^\dag$Assumed parameters:
$H_0=70$ km s$^{-1}$ Mpc$^{-1}$, $\Omega_m=0.28$, and $\Omega\equiv
\Omega_m+\Omega_r+\Omega_\Lambda=1$.}
\end{table*}

A second caveat is that parts of the SED, such as the MIR,
change for different quasar properties. Though the shape
of the MIR is very similar for optically blue and optically red
quasars, there are significant differences between the most
and least optically luminous quasars in their sample.  The
optically luminous quasars are much brighter in the 4 $\mu$m
region than the least optically luminous objects, which is probably
due to physical effects, such as orientation and dust temperature.

A final caveat is that bolometric corrections and bolometric
luminosities determined by summing up all of the observed flux are
in reality line-of-sight values that assume quasars are emitting
isotropically, whereas this is known not to be completely correct.
All in all, computing a bolometric luminosity from an optical
luminosity by assuming a single mean quasar SED may lead
to errors as large as $\sim 50\%$ \cite{R06a}.

These caveats notwithstanding, all of the SEDs constructed in refs. 
\cite{R06a,J06} result in a consistent bolometric correction at $3,000\,\AA$.
Taking the bolometric luminosity to encompass all of the emission
from $100\;\mu$m to 10 keV, $\eta$ at this wavelength ranges
from about 5 to 6 for all of the quasar properties
(see figs.~12 and 13 in ref. \cite{R06a}). In fact, 
in the $3,000\,\AA$ rest frame, the differences
in the composite SEDs for all the quasar sub-classes are relatively
small. This therefore appears to be a robust choice of wavelength
for converting monochromatic luminosity to bolometric luminosity,
because the minimum in this region is due to a relative minimum
in the combination of host galaxy contamination in the near-IR
and dust extinction in the UV. Unfortunately, there does not appear
to be any strong trend between the bolometric correction and color
or luminosity, so it is difficult to know when to apply anything other
than the mean bolometric correction, which we will do throughout
this work. 

The acquisition of some of the data quoted in Table~1 was made possible
by the correlation seen between the Mg II FWHM and $L_{3000}$. 
This constraint is even more interesting in view of its observed 
absence at lower redshifts \cite{F08,S09}, which may be attributed to
the fact that the nearby quasars are accreting at a very wide range of 
sub-Eddington rates \cite{W10}. Thus, the emergence of this 
correlation above $z\sim 6$ is evidence that the distant sources 
may be accreting within a narrower range of Eddington fractions. 
Indeed, these results are consistent with most of the 
high-$z$ quasars accreting near Eddington values.\footnote{The
fact that the tight correlation between  the Mg II FWHM and $L_{3000}$
emerges only for $z> 6$ is one of the principal reasons why the
sample we must use to construct the high-$z$ quasar HD cannot
include AGNs and quasars at lower redshifts.} 

Other (more circumstantial) evidence that the high-$z$ quasars are
accreting at near-Eddington rates is based on the maximum black-hole 
mass observed in the local Universe \cite{S09}. Only a few
black-hole masses exceeding $10^{10}\;M_\odot$ have thus far 
been detected \cite{McConnell12,van12}, even after the peak 
of quasar activity at $1< z< 3$. Yet most high-$z$ quasars accreting 
below $L_{\rm Ed}$ would have to be more massive
than $10^{10}\;M_\odot$ in order to produce the fluxes measured at
Earth. In principle, some of the CFHQS quasars have more moderate
luminosities, so they could be accreting at sub-Eddington rates. But
even in this case \cite{W10}, the lower luminosities
are due to smaller masses (closer to $\sim 10^8\;M_\odot$), rather
than to lower accretion rates.

The conclusion from this meticulous work is that the moderate 
to high-luminosity quasars at $z> 6$ appear to be accreting close 
to their Eddington rate. This may not be true for the fainter sources,
which may be accreting at a broad range of Eddington ratios even
for high redshifts. As a result, there may be a practical limit to
the number of suitable high-$z$ ojbects that are useful as standard
candles. This caveat notwithstanding, there is some evidence
that the mean Eddington ratio does increase with redshift
(see, e.g., figure 19 in ref. \cite{Shen12}).
As we shall see shortly, this is quite useful 
in itself for constructing a Hubble Diagram but, more importantly, 
the transition from sub-Eddington to near-Eddington accretion 
rates across $z\sim 6$, and the accumulation of circumstantial 
evidence we have just described, point to a narrowing in the 
range of Eddington factors above $z\sim 6$, with no further evidence
of evolution in their distribution function $\phi(\lambda_{\rm Ed})$
towards higher redshifts.

The high-$z$ quasars appear 
to be in the exponential buildup of their mass, and have not yet
reached the later phase of quasar activity where the accretion
rate declines to the sub-Eddington values we see locally.
For this principal reason, we suggest that high-$z$ quasars 
with a reasonable determination of their $F_{3000}$ flux
density and Mg II line-widths may therefore be
used as reasonable sources to generate a Hubble
Diagram at $z> 6$, well beyond the reach of Type Ia SNe 
and (probably) also beyond the redshift range where most 
GRBs will be detected.

We should also point out an interesting alternative suggestion
to use super-Eddington accreting quasars as cosmological standards
\cite{Wang13}, based on the realization that photon trapping
\cite{Wyithe12} in some sources affects the total emitted radiation
and results in a saturated luminosity for a given black-hole mass,
which can therefore be used to deduce cosmological distances.
In these black-hole systems, the X-ray emission is linked to
the optical-UV spectrum of the accretion disk, and so in principle,
may be identified through hard X-ray observations. Thus far,
however, the best group of AGNs where such processes have
been studied are narrow line Seyfert 1 galaxies, predominantly
at redshifts $z\lesssim 0.3$. It may be difficult to use this
technique to extrapolate to redshifts $>6$, the principal
aim of this paper.

\section{The High-$z$ Quasar Sample and Hubble Diagram}
The entries shown in Table 1 were obtained by pre-assuming
a $\Lambda$CDM model with concordance parameter values. 
These high-$z$ quasars represent the majority of cases for 
which a reasonable estimate of mass has been made to date. 
However, for obvious reasons, this is not an ideal approach 
to take when attempting to use the high-$z$ quasar HD to 
test competing cosmological expansion scenarios. In principle,
one could recalibrate the data for each assumed model and
then check for consistency {\it a posteriori}. Unfortunately,
this appears to be an essential ingredient with any attempt
at using Type Ia SNe for this purpose, since the data are
themselves characterized by four so-called nuisance parameters
that need to be optimized along with the pre-assumed model
\cite{R98,P98,P99,M12a}. 
Fortunately, we do not need to follow this
procedure here, since the high-$z$ quasar data may be
used without pre-assuming a cosmological expansion, but
only under the (reasonable) assumption that the $z> 6$ 
sources are indeed accreting within a narrow range of 
Eddington factors, presumably centered on a value close 
to one and, most importantly, that their Eddington luminosity
function $\phi(\lambda_{\rm Ed})$  is not changing with redshift.

We will first examine whether the idea of constructing a 
high-$z$ HD can lead to useful results, given the various 
uncertainties associated with the measurements themselves. 
In the expression for $\lambda_{\rm Ed}$, the Eddington rate
is defined from the maximum 
luminosity attainable due to outward radiation pressure acting on 
highly ionized infalling material \cite{M09}. This power 
depends somewhat on the gas composition, but for hydrogen 
plasma is given as $L_{\rm Ed}\approx 1.3\times 10^{38}
(M/M_\odot)$ ergs s$^{-1}$, in terms of the accretor's mass, $M$.
For a more general composition, in which the electron's mean atomic
weight is $\mu_e$, this expression becomes $L_{\rm Ed}\approx 
1.3\times 10^{38}\mu_e (M/M_\odot)$ ergs s$^{-1}$. (For example, 
$\mu_e=2$ when the accreting plasma is pure helium.)

The distribution of Eddington ratios $\lambda_{\rm Ed}$ in quasars is an important
probe into the quasar activity and black-hole growth. Lower redshift
studies \cite{K06,S08} show that the $\lambda_{\rm Ed}$
distribution up to $z=4$ in luminosity and redshift bins is a lognormal
that shifts to higher $\lambda_{\rm Ed}$ and narrows for the higher luminosities.
The most luminous quasars (i.e., $L_{bol}>10^{47}$ ergs s$^{-1}$) at $2<z<3$ 
have a typical $\lambda_{\rm Ed}=0.25$ and dispersion of 0.23 dex \cite{S08}. 
The latest results (see, e.g., figure~6 in ref. \cite{W10}) show that the $\lambda_{\rm Ed}$ 
distribution at $z=6$ can also be approximated by a lognormal, though
here with peak $\lambda_{\rm Ed}=1.07$, and dispersion 0.28 dex. 
The available evidence, though mostly circumstantial, suggests
that the $\lambda_{\rm Ed}$ distribution at higher redshifts remains
centered near one, as we have described above.

\begin{table*}
\center
  \centerline{Table 2. Measured (Dimensionless) Luminosity Distances}\vskip 0.1in
  \begin{tabular}{lccccl}
\hline\hline
&&&&& \\
    Name &  $\quad z\quad$ & FWHM (Mg II) & $F_\nu (3000\,\AA)$ & $\Delta_L$ & Ref. \\
              &  & km s$^{-1}$ & $10^{-29}$ ergs$/$cm$^2/$s$/$Hz&& \\ \hline
&&&&& \\
ULAS J1120+0641& 7.085 &$3800\pm200$&$5.82\pm0.43$&$6.0\pm0.9$&\cite{M11}\\
CFHQS J0210-0456&6.438&$1300\pm350$&$0.67\pm 0.08$&$2.1\pm1.5$&\cite{W10}\\
SDSS J1148+5251& 6.419 &$6000\pm850$ &$12.7\pm 0.20$&$10.1\pm3.1$&\cite{W10}\\
CFHQS J2329-0301&6.417&$2020\pm110$&$1.13\pm 0.13$&$3.8\pm0.7$&\cite{W10}\\
SDSS J1030+0524& 6.310 &$3600\pm100$&$5.64\pm 0.09$&$5.5\pm0.3$&\cite{K07}\\
CFHQS J0050+3445&6.253&$4360\pm270$&$6.42\pm 0.64$&$7.5\pm1.4$&\cite{W10}\\
SDSS J1623+3112& 6.250 &$3600\pm411$&$5.43\pm 0.15$&$5.6\pm1.4$&\cite{J07}\\
SDSS J1048 + 4637&6.198&$3366\pm532$&$12.7\pm 0.10$&$3.2\pm1.1$&\cite{D11}\\
CFHQS J0221-0802&6.161&$3680\pm1500$&$5.45\pm 0.55$&$5.8\pm5.8$&\cite{W10}\\
CFHQS J2229+1457&6.152&$1440\pm330$&$1.02\pm 0.10$&$2.1\pm1.2$&\cite{W10}\\
CFHQS J1509-1749&6.121&$4420\pm130$&$7.54\pm 0.75$&$7.1\pm0.9$&\cite{W10}\\
CFHQS J2100-1715&6.087&$3610\pm420$&$1.69\pm 0.17$&$10.0\pm3.1$&\cite{W10}\\
SDSS J0303-0019& 6.080 &$2300\pm125$&$1.8\pm 0.03$&$3.9\pm0.5$&\cite{K09}\\
SDSS J0353 + 0104&6.072&$3682\pm281$&$6.48\pm 0.3$&$5.3\pm1.0$&\cite{D11}\\
SDSS J0842 + 1218&6.069 &$3931\pm257$&$7.14\pm 0.36$&$5.8\pm1.0$&\cite{D11}\\
SDSS J1630 + 4012&6.058&$3366\pm533$&$4.38\pm 0.9$&$5.4\pm2.8$&\cite{D11}\\
CFHQS J1641+3755&6.047&$1740\pm190$&$2.09\pm 0.23$&$2.1\pm0.6$&\cite{W10}\\
SDSS J1306 + 0356&6.020&$4500\pm160$&$5.91\pm 0.12$&$8.3\pm0.7$&\cite{K07}\\
CFHQS J0055+0146&5.983&$2040\pm280$&$1.12\pm 0.12$&$3.9\pm1.5$&\cite{W10}\\
SDSS J1411+1217&5.950&$2400\pm150$&$9.09\pm 0.18$&$1.9\pm0.3$&\cite{K07}\\ 
&&&&& \\ \hline\hline
  \end{tabular}
\end{table*}

At $z=6$, the peak of the distribution is therefore four times higher than 
for the most luminous quasars at $2<z<3$. Indeed, as we noted in the 
previous section, the typical quasar at $z=6$ appears to be accreting 
right at the Eddington limit, with only a narrow distribution in 
$\lambda_{\rm Ed}$. But the width of this distribution is not 
negligible and an important question that needs to be answered is
whether one may still use the high-$z$ quasars as standard candles 
in spite of this spread. To circumvent this problem, we will
average over $\phi(\lambda_{\rm Ed})$ at each sampled redshift. As
we shall see, the benefit of this method is that it avoids the inevitable 
scatter produced by the spread in individual $\lambda_{\rm Ed}$ values.

Let us now quickly review the newly defined quantities and the
assumptions we have made, while deriving an expression for the 
luminosity distance in terms of the observable parameters. Beginning 
with Equation~(2.2) and the definition of the Eddington luminosity, we
get
\begin{eqnarray}
{L_{\rm Ed}\over 1.3\times 10^{38}\;{\rm ergs}\;{\rm s}^{-1}\;\mu_e}&=&
10^{6.86}\left({\rm FWHM(MgII)\over 1,000\;{\rm km}\;{\rm s}^{-1}}\right)^2 \times\nonumber \\ 
&\null&\left({L_{3000}\over 10^{44}\;{\rm ergs}\;{\rm s}^{-1}}\right)^{1/2}\;,
\end{eqnarray}
where FWHM(MgII) is the Mg II line width at half-maximum, $L_{3000}$ is
the luminosity at rest-frame $3000\;\AA$, and $\mu_e$ is the electron's
mean atomic weight ($=1.17$ for cosmic abundances; see below), which
we assume does not change over the redshift range of interest ($z<1000$).
Thus, in terms of the bolometric factor $\eta\equiv L/L_{3000}$ 
(assumed to have the value $6$) and Eddington factor $\lambda_{\rm Ed}
\equiv L/L_{\rm Ed}$ (presumably of order 1), this becomes
\begin{eqnarray}
{\eta\;L_{3000}\over 1.3\times 10^{38}\;{\rm ergs}\;{\rm s}^{-1}\;\mu_e
\;\lambda_{\rm Ed}}&=&10^{6.86}\left({\rm FWHM(MgII)\over 
1,000\;{\rm km}\;{\rm s}^{-1}}\right)^2 \times \nonumber \\
&\null&\left({L_{3000}\over 10^{44}\;{\rm ergs}\;{\rm s}^{-1}}\right)^{1/2}\;.
\end{eqnarray}
Finally, replacing $L_{3000}$ with the expression
\begin{equation}
L_{\lambda}=4\pi c\,d_L^2{F_\nu\over \lambda}\;,
\end{equation}
where $d_L$ is the luminosity distance and $\nu=c/\lambda$, we arrive at
\begin{eqnarray}
d_L&=&(55.0\;{\rm Glyr})\,\lambda_{\rm Ed}\,\left({\mu_e\over 1.17}\right)
\left({\eta\over 6}\right)^{-1}\left({{\rm FWHM}\over 1,000\,{\rm km}\,
{\rm s}^{-1}}\right)^2\times \nonumber \\
&\null&\qquad\qquad\qquad \left({F_\nu\over 10^{-29}\,{\rm ergs}\,
{\rm cm}^{-2}\,{\rm s}^{-1}\,{\rm Hz}^{-1}}\right)^{-1/2}.
\end{eqnarray}

We will write this equation as
\begin{equation}
d_L(z)=\lambda_{\rm Ed}\,K\Delta_L(z)\;,
\end{equation}
where
\begin{equation}
\Delta_L(z)\equiv\left({{\rm FWHM}\over 1,000\,{\rm km}\,
{\rm s}^{-1}}\right)^2\left({F_\nu\over 10^{-29}\,{\rm ergs}\,
{\rm cm}^{-2}\,{\rm s}^{-1}\,{\rm Hz}^{-1}}\right)^{-1/2}\;,
\end{equation}
and the constant $K$ incorporates all of the other factors appearing in
Equation~(3.4), except for $\lambda_{\rm Ed}$. That is,
\begin{equation}
K\equiv (55.0\;{\rm Glyr})\,\left({\mu_e\over 1.17}\right)\left({\eta\over 6}\right)^{-1}\;.
\end{equation}
In these expressions, the mean molecular weight per electron,
\begin{equation}
\mu_e\approx{2\over 1+X_{\rm H}}\;,
\end{equation}
has been scaled to the value, $1.17$, corresponding to cosmic abundances,
i.e., a fraction $X_{\rm H}\approx 0.7$ of Hydogen by mass.

Expressed in this fashion, $d_L$ and $\Delta_L$ are independent of any cosmological model,
though in principle, all three quantities $\mu_e$, $\eta$, and $\lambda_{\rm Ed}$
may change from quasar to quasar. But since $\mu_e$ depends primarily
on the helium to hydrogen abundance ratio, which changed very little since
big bang nucleosynthesis, and $\eta$ appears to be independent of source
under a wide range of conditions (see previous section), one may reasonably 
expect these two parameters (and therefore $K$) to be nearly constant.

Insofar as $\lambda_{\rm Ed}$ is concerned, the data suggest a compression
of Eddington factors with an average $\langle\lambda_{\rm Ed}\rangle\sim 
1.07$ for $z\sim6-7$. Nonetheless, one cannot ignore the fact that the 
distribution $\phi(\lambda_{\rm Ed})$, normalized such that
\begin{equation}
\int \phi(\lambda_{\rm Ed})\;d\lambda_{\rm Ed}=1\;,
\end{equation}
is lognormal with a dispersion 0.28 dex. So even though the average
$\langle\lambda_{\rm Ed}\rangle$ is close to 1 and apparently independent
of redshift for $z> 6$, the bolometric luminosity of individual quasars may 
deviate from their Eddington limit by factors of $\sim 2$. Thus, instead of
constructing the HD from individual sources, we can avoid the consequent
scatter produced by this spread in $\lambda_{\rm Ed}$  by simply averaging
over the Eddington factor at each sampled redshift.

In principle, $\phi$ may also depend on $z$, and the data do show
that this is the case for $z\le 6$, as we have noted earlier. But since
$\langle\lambda_{\rm Ed}\rangle$ levels off at $\sim 1$ for
$z\sim 6-7$, we will assume that in this range $\phi$ is only a function
of $\lambda_{\rm Ed}$. Within a small redshift bin $\Delta z=z_2-z_1$ at 
$z=(z_1+z_2)/2$, we may then write
\begin{equation}
\langle d_L(z)\rangle=K\,\langle\lambda_{\rm Ed}\rangle\,\langle\Delta_L(z)\rangle_{\rm sample}\;,
\end{equation}
where
\begin{equation}
\langle\lambda_{\rm Ed}\rangle\equiv \int d\lambda_{\rm Ed}\;\phi(\lambda_{\rm Ed})\,
\lambda_{\rm Ed}\;,
\end{equation}
and $\langle\Delta_L(z)\rangle_{\rm sample}$ is the sample average in that bin.
The advantage of using these averages over individual sources is rather clear.
If all of the assumptions and inferences we have made leading up to this equation
are valid, the expectation is that both $K$ and $\langle\lambda_{\rm Ed}\rangle$
($\sim 1$) should be constant. Thus, by finding the sample average $\langle\Delta_L(z)
\rangle_{\rm sample}$ in each bin $\Delta z$, one may infer the average luminosity distance 
$\langle d_L(z)\rangle$, and use it to construct the high-$z$ quasar HD to test
competing cosmological models.

\begin{table}
\center
\centerline{Table 3. Binned Sample}\vskip 0.1in
  \begin{tabular}{lcc}
\hline\hline
&& \\
    Redshift &  $\langle \Delta_L\rangle$ & $\sigma$ \\ 
&& \\ \hline
&& \\
$5.95-6.05$&4.1&2.6\\
$6.05-6.15$&6.3&1.9\\
$6.15-6.25$&3.7&1.6\\
$6.25-6.35$&6.2&0.9\\
$6.35-6.45$&5.3&3.4\\
$7.05-7.15$&6.0&0.9$^\dag$\\ 
&& \\ \hline\hline
  \end{tabular}
{\vskip 0.1in $^\dag$Calculated from the standard deviation of \\ the individual
source (ULAS J1120+0641) \\ in this bin. For comparison, we also cons-\\ ider an
outcome based on the use of a\\ sample-averaged standard deviation\\ (i.e., $\sigma=1.9$)
for the $z=7.05-7.15$ bin.}
\end{table}

The individual dimensionless source distances $\Delta_L$ measured in this way 
are listed in Table 2, together with the flux densities $F_\nu$ and
Mg II line-widths used to calculate them. With the catalog 
of 20 sources from Table 1, a reasonable bin size is $\Delta z=0.1$, 
which one may use to calculate the sample-averaged values 
$\langle \Delta_L(z)\rangle$ quoted in Table 3. The third 
column shows the population standard deviation 
for each bin, except for $z=7.05-7.15$. Since this bin contains only 
one source, the quoted standard deviation is simply that of 
ULAS J1120+0641 itself. However, in order to ensure that this
particular source does not unduly influence the optimization of
the fits relative to the other bins, in our analysis below we will
also consider an outcome based on the use of a sample-averaged
standard deviation for the highest redshift bin.

The binned data in Table 3 are plotted in figure~1, together with the
theoretical fits we will discuss in the next section. A quick estimate of the
luminosity distance calculated from Equation~(3.5), using the scaling in
Equation~(3.7), shows that with these values our ``measured" distance
$d_L(z)$ appears to over-estimate the luminosity distance one would
expect in the standard model (see Table 1) by $\sim 20-40\%$. There
are several possible reasons for this. One of them is that we have
carefully included the dependence of $L_{\rm Ed}$ on $\mu_e$, which
was ignored in earlier applications \cite{W10}. Unfortunately,
we cannot know for sure what abundances characterize the medium
surrounding these sources, but the introduction of $\mu_e$ into
these expressions produces at least a $\sim 17\%$ difference from
previously calculated $\lambda_{\rm Ed}$ values. Secondly, as we have
alluded to previously, there appears to be at least a $\sim 20\%$ uncertainty
in the value of $\eta$. So, for example, if we were to use the scaling
$K=(40.3\;{\rm Glyr})\,(\mu_e/1.0)(\eta/7)^{-1}$, instead of Equation~(3.7),
the distances measured with Equation~(3.5) would be right in line with
the luminosity distances quoted for the standard model in Table 1.

\begin{figure}
\begin{center}
\includegraphics[width=0.7\linewidth]{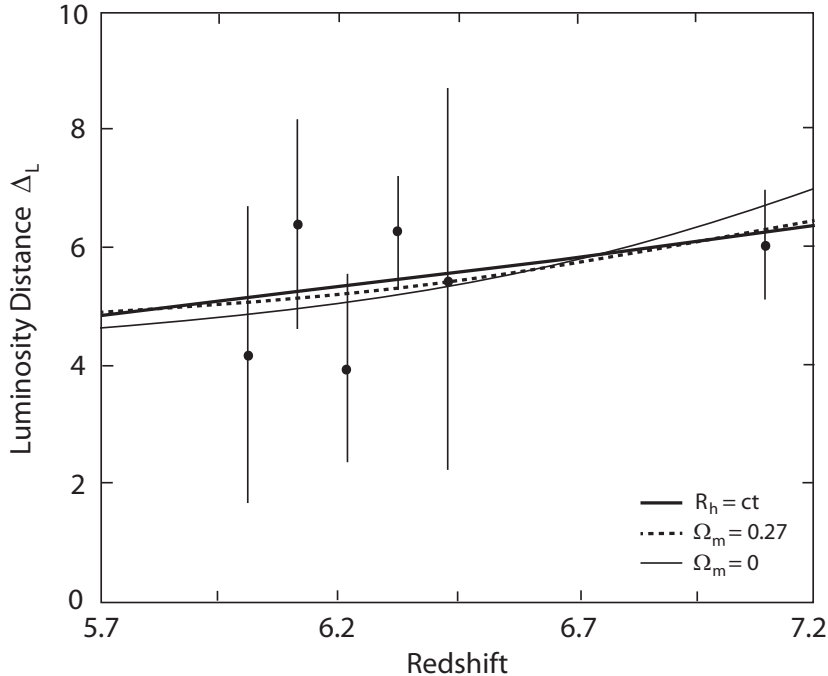}
\end{center}
\caption{Sample-averaged (dimensionless) luminosity distance 
$\langle \Delta_L\rangle$ from Table 3. The solid curve gives the 
corresponding luminosity distance $d_L^{R_{\rm h}=ct}$ in the 
$R_{\rm h}=ct$ Universe, while the best fit $\Lambda$CDM model,
with $\Omega_m=0.27$ and $w_{de}=-1$, is shown as a dashed line.
For comparison, we also show a $\Lambda$CDM model with $\Omega_m=0$
(thin solid line), to highlight the dependence of these redshift-distance
relationships on the parameters. The curves shown here all correspond 
to the values $K=40.3$ Glyr, $\langle \lambda_{\rm Ed}\rangle=1$, and $H_0=70.0$
km s$^{-1}$ Mpc$^{-1}$. To produce these fits, we need 3 
parameters for $\Lambda$CDM and 1 for $R_{\rm h}=ct$. Therefore, 
the reduced $\chi^2_{\rm dof}$ for these curves is 0.76 for the 
optimized $\Lambda$CDM, 0.88 for $\Lambda$CDM with $\Omega_m=0$, 
and 0.48 for $R_{\rm h}=ct$.}
\end{figure}

Fortunately, these uncertainties do not affect the shape of our
$\langle d_L(z)\rangle$ curve calculated from Equation~(3.10). All 
Friedmann-Robertson-Walker metrics have a luminosity distance 
proportional to $c/H$ (see below), so our current imprecise knowledge
of these factors directly affects the value of the Hubble constant that 
we could infer from fits to the high-$z$ quasar data. Since the
overall uncertainty in $K$ appears to be $\sim 20-40\%$, it
therefore does not make sense to worry about optimizing the value
of $H$ in these fits, since measurements of $H$ using other techniques
are much more reliable. Thus, until $\mu_e$ and $\eta$ are known more
precisely, we will compare how well competing cosmologies do in fitting
the high-$z$ quasar HD by concentrating solely on the shape of the
$\langle d_L(z)\rangle$ or, equivalently, the $\langle\Delta_L(z)\rangle$,
distributions in figure~1. But to illustrate how the value of $H_0$
would have impacted the fits to the quasar data, we show in 
figures~2 and 3 the luminosity distance versus redshift for the 
data in figure~1, and three curves: (a) $H_0=60$  km s$^{-1}$ Mpc$^{-1}$,
(b) $H_0=69.32$  km s$^{-1}$ Mpc$^{-1}$ (the Planck best-fit value; 
see Ade et al. \cite{Ade13}), and (c) $H_0=80$  km s$^{-1}$ Mpc$^{-1}$.

\section{Theoretical Fits to the High-$z$ Quasar HD}
To demonstrate the future potential for using the high-$z$ quasar
HD in order to distinguish between competing cosmologies, we will
here compare the entries in Table 3 with two different
expansion scenarios: $\Lambda$CDM (with its three free parameters,
$H_0$, $\Omega_m$, and the dark-energy equation of state
$w_{de}$) and the $R_{\rm h}=ct$ Universe, which has only one
free parameter, the Hubble constant $H_0$ (though $H_0$ will not
be optimized here for either cosmology; see \S3 above). If we choose $w_{de}=-1$
(thus reducing the number of free parameters to 2), it is not difficult 
to show that in $\Lambda$CDM the expected luminosity distance is \cite{M12a}
\begin{equation}
d_L^{\Lambda{\rm CDM}}={c\over H_0}(1+z)\int_{1\over 1+z}^1
{du\over\sqrt{\Omega_r+u\Omega_m+u^4\Omega_\Lambda}}\;,
\end{equation}
in terms of the scaled radiation ($\Omega_r$),
matter ($\Omega_m$), and dark-energy ($\Omega_\Lambda$) densities.

\begin{figure}
\begin{center}
\includegraphics[width=0.7\linewidth]{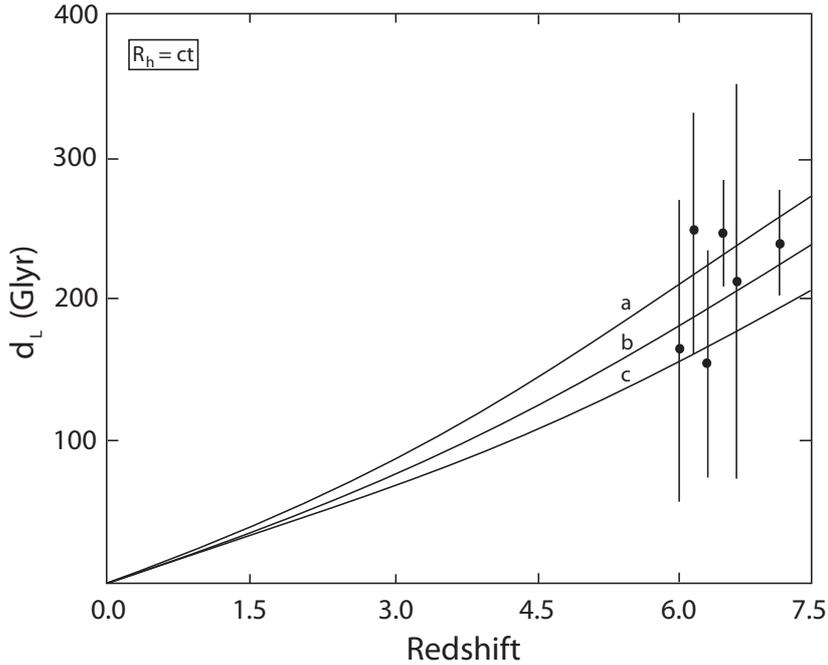}
\end{center}
\caption{The luminosity distance versus redshift for the
same data shown in figure~1. The curves illustrate the
dependence of the fit on the Hubble constant and
are for the $R_{\rm h}=ct$ Universe with three different
values of $H_0$: (a) $60$  km s$^{-1}$ Mpc$^{-1}$,
(b) $69.32$  km s$^{-1}$ Mpc$^{-1}$ (the Planck best-fit value; 
see Ade et al. 2013), and (c) $80$  km s$^{-1}$ Mpc$^{-1}$.
The other parameters are $K=40.3$ Glyr and $\langle\lambda_{\rm Ed}
\rangle=1$.}
\end{figure}

The $R_{\rm h}=ct$ cosmology is still not widely known, wo we will begin 
by introducing some of its principal features. One way of looking at the 
expansion of the Universe is to guess its constituents and their equations 
of state and then solve the dynamics equations to determine the expansion 
rate as a function of time. This is the approach taken by $\Lambda$CDM. A 
second---though not mutually exclusive---way is to use symmetry 
arguments and our knowledge of the properties of a gravitational 
horizon in general relativity (GR) to determine the spacetime curvature, 
and thereby the expansion rate, strictly from just the value of the total 
energy density $\rho$ and the implied geometry, without necessarily 
having to worry about the specifics of the constituents that make up 
the density itself. This is the approach adopted by $R_{\rm h}=ct$. 
The constituents of the Universe must then partition themselves in 
such a way as to satisfy that expansion rate. In other words, what 
matters is $\rho$ and the overall equation of state $p=w\rho$, in 
terms of the total pressure $p$ and total energy density $\rho$. In
$R_{\rm h}=ct$, it is the aforementioned symmetries and other
constraints from GR that uniquely fix $w$ to have the value
$-1/3$ \cite{M07,M12b}. 

The $R_{\rm h}=ct$ Universe is a Friedmann-Robertson-Walker (FRW) 
cosmology in which Weyl's postulate takes on a more important role 
than has been considered before. Most workers assume that 
Weyl's postulate is already incorporated into all FRW metrics, but actually 
it is only partially incorporated. Simply stated, Weyl's postulate says that any 
proper distance $R(t)$ must be the product of a universal expansion 
factor $a(t)$ and an unchanging co-moving radius $r$, such that
$R(t)=a(t)r$. The conventional way of writing an FRW metric
adopts this coordinate definition, along with the cosmic time $t$, 
which is actually the observer's proper time at his/her location. 
But what is often overlooked is the fact that the gravitational radius, 
$R_{\rm h}\equiv c/H$, which has the same definition as the Schwarzschild 
radius, and actually coincides with the better known Hubble radius, is in
fact itself a proper distance too \cite{MA09}. And when one 
forces this radius to comply with Weyl's postulate, there is only one possible 
choice for $a(t)$, i.e., $a(t)=(t/t_0)$, where $t_0$ is the current age of the
Universe. This also leads to the result that the gravitational radius must
be receding from us at speed $c$, which is in fact how the Hubble radius 
was defined in the first place, even before it was recognized as another
manifestation of the gravitational horizon. 

The fact that $p=-\rho/3$ in $R_{\rm h}=ct$ means that quantities, such 
as the luminosity distance and the redshift-dependent Hubble constant
$H(z)$, take on very simple, analytical forms \cite{M12a,MM13}:
\begin{equation}
d_L^{R_{\rm h}=ct}={c\over H_0}(1+z)\ln(1+z)\;,
\end{equation}
and
\begin{equation}
H(z)=H_0(1+z)\;.
\end{equation}
Yet even though these functional forms are quite different from 
their $\Lambda$CDM counterparts, in the end, regardless of how 
$\Lambda$CDM and $R_{\rm h}=ct$ handle $\rho$ and $p$, 
they must both account for the same cosmological data. And 
there is now growing evidence that $\Lambda$CDM functions 
as a reasonable approximation to $R_{\rm h}=ct$ in some 
redshift ranges, but apparently not in others, as discussed in
the introduction. Interestingly, we will find that here too, with
the high-$z$ quasar HD, the optimized $\Lambda$CDM model
that best fits the data comes as close as its parametrization
allows it to the $R_{\rm h}=ct$ curve.

\begin{figure}
\begin{center}
\includegraphics[width=0.7\linewidth]{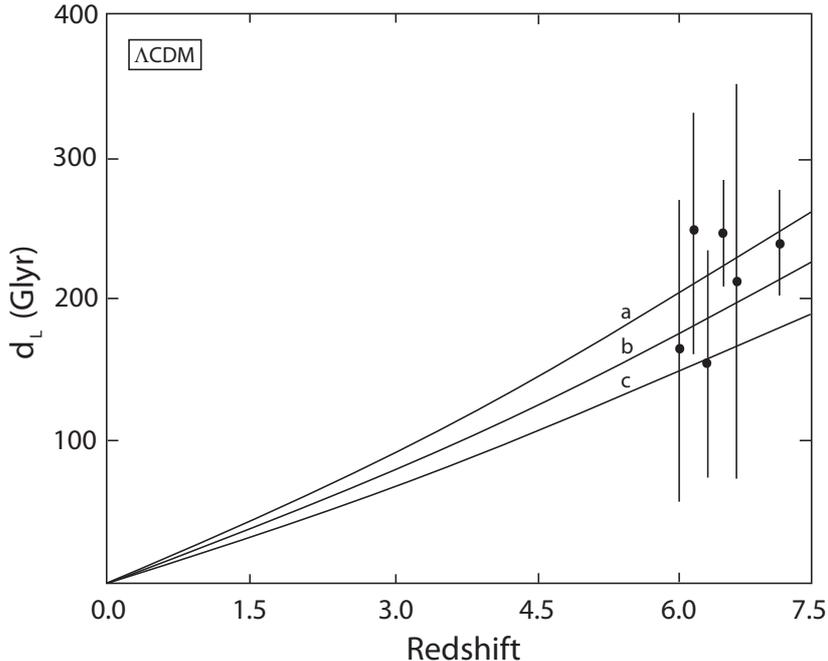}
\end{center}
\caption{Same as figure~2, except now for $\Lambda$CDM.
In addition to the parameters $K=40.3$ Glyr and $\langle\lambda_{\rm Ed}
\rangle=1$, these curves also assume $\Omega_m=0.29$ (again from
the Planck best fit), and a dark-energy equation-of-state $w_{de}\equiv
w_\Lambda=-1$.}
\end{figure}

The theoretical curves that best fit the data are shown in
figure~1, for both the $R_{\rm h}=ct$ Universe (solid, thick
line) and $\Lambda$CDM (dashed line). To gauge the dependence
of these results on the parameters, we also show the curve
corresponding to $\Lambda$CDM with $\Omega_m=0$.
(The more general dependence of the $\Lambda$CDM fit
on the value of $\Omega_m$ is shown in figure~4.)
With only one free parameter, the $\chi^2_{\rm dof}$ for
$R_{\rm h}=ct$ is 0.48, compared with 0.76 and 0.88
for the $\Lambda$CDM fits. Based solely on their $\chi^2$-values, 
one would therefore conclude that all three models
provide reasonable fits to the high-$z$ quasar HD. However, model
selection tools strongly favor models with fewer degrees of
freedom, so the likelihood of any of these models being closest
to the correct cosmology is different for the three cases (see 
\S5 below).

\begin{figure}
\begin{center}
\includegraphics[width=0.7\linewidth]{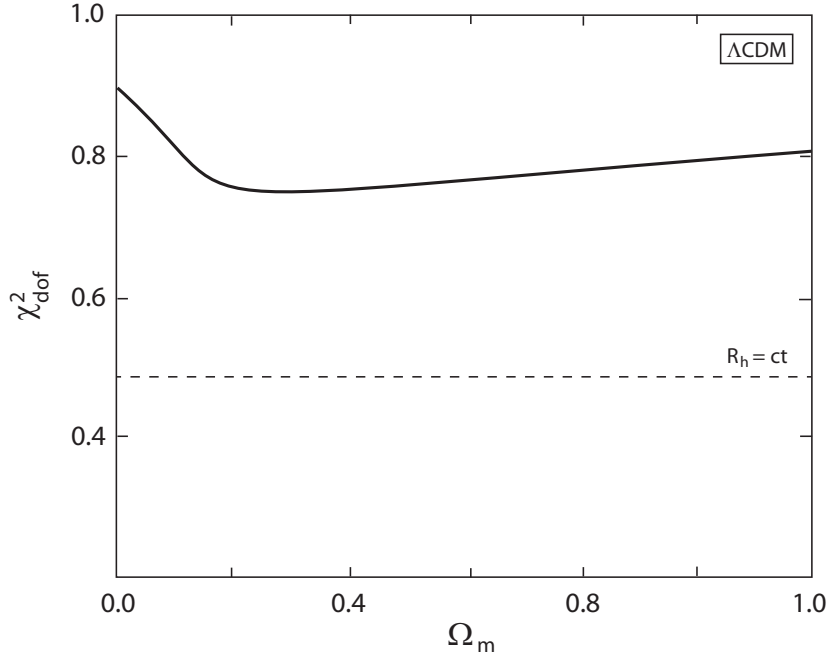}
\end{center}
\caption{Reduced $\chi^2$ for $\Lambda$CDM fits to the data shown
in figure~1, as a function of $\Omega_m$. The minimum $\chi^2_{\rm dof}$ is
realized for $\Omega_m=0.27$, which produces a luminosity distance
curve over this redshift range essentially identical to that in the
$R_{\rm h}=ct$ Universe (see figure~1). For reference, the dashed line
shows the reduced $\chi^2_{\rm dof}$ for the $R_{\rm h}=ct$ Universe
which, however, does not depend on $\Omega_m$.}
\end{figure}

Note that even though Equations~(4.1) and (4.2) could have produced 
dramatically different results (e.g., depending on the choice of 
$\Omega_m$), the best fit $\Lambda$CDM model has parameter values that
bring it closest to the $R_{\rm h}=ct$ Universe. This is the same 
phenomenon that emerged from fits to the Type Ia SNe data \cite{M12a}, 
and to the gamma-ray burst Hubble diagram \cite{W13}, where the 
distance versus redshift relationship produced by the best fit 
$\Lambda$CDM model appears to be relaxing to that expected in 
the $R_{\rm h}=ct$ cosmology.

Though the number of sources used here is still rather small, 
it is already quite evident from these figures that eventually
the catalog of high-$z$ quasars with measured
Mg II line-widths and flux densities at $3000\;\AA$ will be large
enough to significantly reduce the scatter reflected in the
standard deviations listed in Table 3. Much work still 
needs to be done in assembling a high-quality sample of $z> 6$ quasars 
for this kind of study, but these results already suggest that the effort 
will be worthwhile. We notice, in particular, that ULAS J1120+0641 
(at $z=7.085$) fits the theoretical curves remarkably well, confirming 
our suspicion that quasars tend to accrete at a rate closer to the
Eddington value the higher their redshift. 

The importance of this source in anchoring the fits shown in figure~1 
is quite evident, for it provides a significant stretching in the range of
sampled redshifts. But suppose that instead of assigning a standard
deviation of $0.9$ to the highest redshift bin, we use the sample-averaged
value of $2.1$. How would the fits shown in figure~1 be affected
by this change? As it turns out, the optimized parameter values remain the
same, though the reduced $\chi^2_{\rm dof}$'s change slightly. For the
$R_{\rm h}=ct$ Universe, we would now have $\chi^2_{\rm dof}=0.44$
(instead of $0.48$), while for $\Lambda$CDM the corresponding value 
associated with $\Omega_m=0.27$ is $0.73$ (instead of $0.76$). In other 
words, the reason ULAS J1120+0641 has such a large influence on the 
results is not only because of its relatively small error bar, but primarily 
because of its much higher redshift compared to the
other sources listed in Table 3.

\section{Discussion and Conclusions}
In the past decade, over 50 quasars have been discovered 
at $z> 6$ with the help of dedicated programs, such as
the SDSS and CFHQS. Two recent
developments have made it possible for us to start thinking
about using these powerful sources to construct a Hubble
Diagram well beyond the redshift range ($z< 1.8$) accessible
to Type Ia SNe, and even GRBs, most of which are expected to
be discovered at $0<z< 6$. High-$z$ quasars therefore offer
us a unique opportunity of studying the expansion of the Universe
in its very important early epoch, where there appears to be
a paucity of other possible standard candles.

As we have seen, the hypothesis that high-$z$ quasars accrete at close to
their Eddington rate and, especially, that their distribution in Eddington factors 
is the smallest of any redshift range sampled thus far, plus the apparent 
correlation between the line widths in their broad-line region and their optical/UV
luminosity, allows to us to use their sample-averaged Eddington luminosity as a 
standard candle. We have highlighted the inference that, because the actual 
$\lambda_{\rm Ed}$ distribution of these sources
has an observed finite width (roughly 0.28 dex), it is not possible
to use individual sources to construct the HD without introducing some
contamination by quasars with $\lambda_{\rm Ed}> 2$ or 
$\lambda_{\rm ED}< 0.4$. Nonetheless, we have also demonstrated
that as long as the Eddington-factor distribution $\phi(\lambda_{\rm Ed})$
is nearly constant over the redshift range $z\sim 6-8$, we can
use sample-averaged estimates of the luminosity distance to
test competing models. With the procedure we have
described in this paper, we expect that an extended high-quality 
survey will thus permit us to probe the history of the universe 
over the first 1--2 Gyr of its expansion.

We have also seen how crucial it is to extend the quasar redshift
range beyond 7, which is necessary to provide sufficient leverage
when fitting theoretical luminosity distances to the data. In this
regard, the work we have reported here would not have been
feasible without the recent discovery of ULAS J1120+0641 at
$z=7.085$. Quite remarkably, the inferred luminosity distance
to this source matches the best-fit curves rather well.
This may be somewhat fortuitous, but may also be
a confirmation of the expectation that quasars accrete
closer to their Eddington rate, the higher their redshift,
thus affirming our conclusion that the Eddington-factor
distribution $\phi(\lambda_{\rm Ed})$ probably does remain
narrow and constant towards higher redshifts.
Clearly, every effort should be expended to acquire
additional quasars at $z> 7$. 

An interesting alternative proposal to use Active Galactic 
Nuclei (AGNs) for cosmological distance measurements
was made recently \cite{W11}. This idea
is also based on the observed relationship between the
luminosity of type 1 AGNs and the sizes of their 
broad-line regions (see Eq.~1), but using the actual
observed time delay $\tau$ between the response of the
flux in the broad lines to variations in the luminosity
of the central source in order to calculate the radius,
$R$, of the BLR. With this approach, the observable
quantity $\tau/\sqrt{F}$, where $F$ is the
AGN continuum flux, is then a measure of the luminosity
distance to the source. 

At least for the forseeable
future, however, this method probably cannot be
used to construct the quasar HD at the high redshifts
we are considering here. The problem is that extending
the catalog to high redshifts requires substantially 
longer temporal baselines because (1) redshift increases
the observed-frame lags due to time dilation effects, and
(2) at higher redshifts we observe more luminous
AGNs, which have larger BLRs and hence larger
rest-frame lags. For example, Watson et al.
estimate a H$\beta$ observed lag of $\sim 2$ years
at $z\sim 2$. The lag reaches close to a decade
at $z\sim 2.5$, making this the practical redshift
limit for obtaining lags with H$\beta$. Strong UV
lines, such as C IV 1549 $\AA$, can do better because
the BLR is ionization-stratified, so these higher
excitation lines are emitted closer to the central
source. Thus, C IV lags could in principle be
measurable for objects up to $z\sim 4$, but almost
certainly not beyond $z\sim 5-6$. The approach
we have described in this paper therefore has the
unique potential of extending cosmological distance
measurements to redshifts well beyond even this
alternative use of high-$z$ AGNs.

The construction of a high-$z$ quasar HD has allowed
us to continue our comparison of $\Lambda$CDM with the
$R_{\rm h}=ct$ Universe, which has so far been superior
to the former in being able to account for several hitherto
inexplicable coincidences and observations that appear to
be at odds with the predictions of the standard model.
For example, we have recently shown \cite{M13} that,
whereas $\Lambda$CDM cannot explain the implied
early appearance of $10^9\;M_\odot$ supermassive
black holes without invoking anomalously high accretion
rates or the creation of exotically massive seeds, neither
of which is seen in the local universe, in the $R_{\rm h}=ct$
Universe, $5-20\;M_\odot$ seeds produced from the deaths
of Pop II and III stars at $z< 15$ could have easily grown
to $M\sim 10^9\;M_\odot$ by $z\sim 6$, merely by accreting
at the standard Eddington rate. The recent observations 
we have discussed in this paper have compounded the problem
for $\Lambda$CDM by demonstrating that in fact all of the 
high-$z$ quasars appear to be accreting at or below this rate, 
commensurate with the expectations in the $R_{\rm h}=ct$ 
cosmology.

In this paper, we have found that both $R_{\rm h}=ct$ and
$\Lambda$CDM fit the current catalog of high-$z$ quasars quite 
well. It is noteworthy that the $\Lambda$CDM fit is optimized
for the parameter value $\Omega_m=0.27$, consistent with
expectations from the concordance model. As we have demonstrated
before \cite{M12a}, this choice of parameters results in a luminosity 
distance versus redshift relation virtually identical to that of $R_{\rm h}
=ct$ (compare the $\Omega_m=0.27$ curve in figure~1 with the 
curve for $R_{\rm h}=ct$).  This appears to be further evidence 
that the parameters in $\Lambda$CDM allow it to relax to the 
$R_{\rm h}=ct$ expansion profile when fitting the data.

Having said this, the process of model selection does take into
account the number of unrestricted parameters used in the optimization
process, and in this regard, the result of our analysis in this paper tends 
to favor $R_{\rm h}=ct$ over $\Lambda$CDM.
In a companion paper \cite{MM13}, we examined
how the redshift dependence of the Hubble constant $H(z)$ predicted
by the $R_{\rm h}=ct$ and $\Lambda$CDM cosmologies compares
with the cosmic chronometer data. These measurements sample
a much lower redshift range (typically $z< 2$) \cite{Mo12}. 
In that work, we discussed at length
how one may use state-of-the art statistical tools to 
select the model most likely to be correct in accounting
for the data. Though we will not reproduce that extensive
discussion here, we do point out that to compare the evidence 
for and against competing models, such as models of the 
distance--redshift relationship, the use of the Akaike Information
Criterion (AIC) is now common in cosmology \cite{L04,L07,T12}. 
The AIC can be viewed as an enhanced 
``goodness of fit'' criterion, which extends the usual~$\chi^2$ 
criterion by taking account of the number of parameters in each 
model. 

The AIC provides the relative ranks of two or more competing models, 
and also a numerical measure of confidence that each model is the best.  
These confidences are analogous to likelihoods or posterior probabilities 
in traditional statistical inference, but unlike traditional inference
methods, the AIC can be applied to models that are not ``nested,''
such as we have here. The AIC for the fitted model is given by
\begin{equation}
{\rm AIC}=\chi^2+2n\;,
\end{equation}
where $n$ is the number of unknown parameters. Then, a quantitative 
ranking of models 1 and 2 can be computed as follows. If ${\rm AIC}_\alpha$ 
comes from model $\alpha$, the unnormalized confidence that this model
is true is the ``Akaike weight'' $\exp(-{\rm AIC}_\alpha/2)$.  Informally,
model $\alpha$ has likelihood
\begin{equation}
P(\alpha)=
\frac{\exp(-{\rm AIC}_\alpha/2)}
{\exp(-{\rm AIC}_1/2)+\exp(-{\rm AIC}_2/2)}
\end{equation}
of being the correct choice. From the analysis we have carried out in
this paper, the Akaike Information Criterion has the value
${\rm AIC}_1\approx 4.4$ for $R_{\rm h}=ct$, compared with
${\rm AIC}_2\approx 8.3$ for $\Lambda$CDM. Therefore, the probability
of $R_{\rm h}=ct$ being the correct choice is $\approx 87\%$,
while $\Lambda$CDM would be selected with a relative confidence 
level of only $\approx 13\%$. Two other commonly used model
selection criteria are the Kullback Information Criterion (KIC)
\cite{C04} and the Bayes Information Criterion (BIC) \cite{S78},
defined as ${\rm KIC}=\chi^2+3n$ and ${\rm BIC}=\chi^2+(\ln N)n$, where
$N$ is the number of measurements. The result of our fitting
shows that both the KIC and BIC favor $R_{\rm h}=ct$ over 
$\Lambda$CDM by a ratio of about $85-95\%$ to $5-15\%$. Thus,
all three of these statistical tools confirm each other's outcome---that 
the high-$z$ quasars reveal a preference for $R_{\rm h}=ct$ 
over $\Lambda$CDM.

Of course, no one would suggest yet that the probabilities 
we have calculated here are sufficient on their own to clinch 
the case for $R_{\rm h}=ct$, but they do reinforce the 
conclusion arrived at elsewhere, that
this cosmology likely provides the correct expansion history
for the Universe, while $\Lambda$CDM is a parameter-driven 
approximation to it.

Even though the Hubble Diagram we have constructed here is 
limited by relatively large uncertainties (due primarily
to the still small sample), the results we have reported in 
this paper do suggest that high-$z$ quasars may eventually 
yield information on the luminosity distance at $z> 6$ with 
sufficient precision for us to carry out a meaningful 
examination of the Universe's early expansion history, 
complementing the highly detailed study already underway 
(with both Type Ia SNe, GRBs, and cosmic chronometers) 
at lower redshifts. 

\acknowledgments
I am grateful to the many workers who spent an extraordinary amount of 
effort and time accumulating the data summarized in Tables 1--3. I am
also very thankful for the thoughtful and constructive comments of the
anonymous referee, resulting in a significantly improved
manuscript. I acknowledge Amherst College for its support through a 
John Woodruff Simpson Lectureship. This work was partially carried 
out at the Purple Mountain Observatory in Nanjing, China.

\end{document}